\documentclass[twocolumn,showpacs,preprintnumbers,amsmath,amssymb,superscriptaddress,prl]{revtex4}

\usepackage{graphicx}
\usepackage{dcolumn}
\usepackage{bm}

\begin{document}

\preprint{APS/123-QED}

\title{Synthesis and Analysis of Entangled Photonic Qubits in Spatial-Parity Space}

\author{Timothy Yarnall}

\affiliation{Quantum Imaging Laboratory, Departments of Electrical
\& Computer Engineering and Physics, Boston University, Boston,
Massachusetts 02215-2421, USA}

\author{Ayman F. Abouraddy}

\affiliation{Research Laboratory of Electronics, Massachusetts
Institute of Technology, Cambridge, Massachusetts 02139-4307, USA}

\author{Bahaa E. A. Saleh}
\author{Malvin C. Teich}
\homepage{http://www.bu.edu/qil}
\affiliation{Quantum Imaging
Laboratory, Departments of Electrical \& Computer Engineering and
Physics, Boston University, Boston, Massachusetts 02215-2421, USA}

\date{\today}

\begin{abstract}
We present the novel embodiment of a photonic qubit that makes use
of one continuous spatial degree of freedom of a single photon and
relies on the the \textit{parity} of the photon's transverse spatial
distribution. Using optical spontaneous parametric downconversion to
produce photon pairs, we demonstrate the controlled generation of
entangled-photon states in this new space. Specifically, two Bell
states, and a continuum of their superpositions, are generated by
simple manipulation of a classical parameter, the optical-pump
spatial parity, and not by manipulation of the entangled photons
themselves. An interferometric device, isomorphic in action to a
polarizing beam splitter, projects the spatial-parity states onto an
even--odd basis.  This new physical realization of photonic qubits
could be used as a foundation for future experiments in quantum
information processing.
\end{abstract}


\maketitle

The generation of entangled states is the cornerstone of
experimental quantum information science \cite{Bouwmeester00Book}.
Photonic entangled systems are of particular interest because of
their importance for quantum cryptography \cite{Gisin02RMP} and
quantum computation with linear optics \cite{Kok07RMP}. The physical
realization of photonic states has hitherto taken the form of
\emph{discrete} degrees of freedom (predominantly in the form of
qubits) such as polarization \cite{Kwiat95PRL} and time--energy bins
\cite{Gisin02PRA}. While the \emph{continuous} spatial degrees of
freedom of entangled photons have been the subject of considerable
interest \cite{Strekalov95PRL}, this feature has heretofore been
only partially harnessed for quantum information processing. One
approach to endowing the spatial degrees of freedom of a photon with
a qubit or a qudit structure is a discretization of the spatial
domain by making use of slits or pinholes \cite{Fiorentino04PRL}.
Another method begins by adopting either the Hermite--Gaussian
\cite{walborn:053812} or Laguerre--Gaussian modes \cite{Mair01Nat}
as a basis for describing a photon's \emph{two}-dimensional
transverse distribution. Typically only two of the initially
infinite number of modes are retained, via filtering or
post-selection, to serve as qubit levels, resulting in a
\emph{truncation} of the Hilbert space. Another approach relies on
so-called \textit{pseudo-spin} operators that have been studied
using photon-number Fock states \cite{Chen02PRL}; however, the
experimental realization of the proposed schemes has not been
forthcoming, undoubtedly due to the difficulty of preparing and
manipulating Fock states \cite{Brukner03PRA}. The pseudo-spin
approach relies on \emph{mapping} (not filtering or truncating) a
Hilbert space associated with a continuous variable onto a discrete
smaller-dimensional space, in particular a two-dimensional (2D) one
\cite{Chen02PRL}, to achieve a qubit structure \emph{without
truncating} the initial Hilbert space.

In this Letter, we present a new physical embodiment of photonic
\textit{qubits} that makes use of a \emph{one}-dimensional (1D)
continuous spatial degree of freedom of single photons that is
readily implemented experimentally without discretization or
truncation. We generate entangled-photonic qubits in this new
Hilbert space using the accessible process of optical spontaneous
parametric downconversion (SPDC) \textit{without} spatial filtering.
Each qubit is encoded in the spatial parity (even--odd) of the
photon's \emph{one}-dimensional transverse modes and is a
realization of the pseudo-spin approach in the spatial domain. The
mathematical underpinning of this approach relies on the isomorphism
between the \textit{single-mode multi-photon} quantization and the
\textit{single-photon multi-mode} quantization of the
electromagnetic field that has recently led to the concept of parity
entanglement, as identified theoretically in
Ref.~\cite{Abouraddy07PRA}. The infinite-dimensional Hilbert space
of \emph{one} spatial degree of freedom for each photon is thereby
mapped onto a 2D Hilbert space describing its spatial-parity qubit.
It is notable that an additional qubit may be encoded in an
identical manner on the orthogonal transverse dimension because $x$
and $y$ are on equal footing as spatial coordinates. This creates a
sharp distinction from approaches based on orbital angular momentum,
which rely on an angular degree of freedom in a polar coordinate
system; encoding an additional qubit in the radial variable would
require a radically different approach.

\begin{figure*}[hb]
\includegraphics[width=7.0in,height=3.9in]{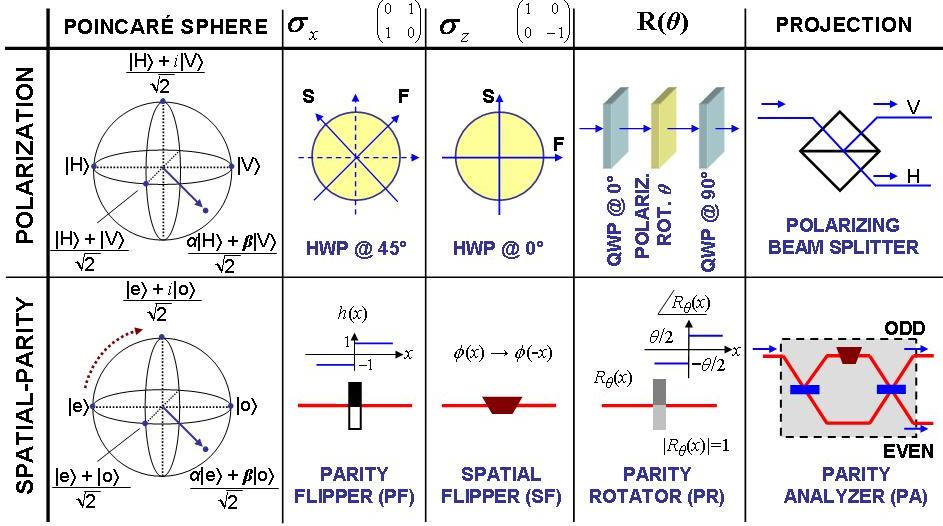}
\caption{(Color online) Comparison of polarization photonic qubits
and operations (first row) and their counterparts in 1D spatial
parity (second row). HWP: half-wave plate; QWP: quarter-wave plate;
H, V: horizontal and vertical polarization; S, F: slow and fast axes
of the wave plate; $\bf{R}(\theta)$: rotation operator;
$\sigma_{x}$, $\sigma_{z}$: Pauli operators.} \label{Comparison}
\end{figure*}

We begin by discussing the construction of operators on the
spatial-parity space, which are strikingly simple to implement, thus
making this approach an attractive alternative to other physical
realizations of photonic qubits. This is highlighted in
Fig.~\ref{Comparison}, where comparison is drawn between the
familiar devices that manipulate polarization qubits, as an
archetypical realization of a photonic qubit, and their isomorphic
counterparts in 1D spatial-parity space (see
Ref.~\cite{Abouraddy07PRA} for a detailed analysis). The 2D manifold
of pure states of polarization (spatial parity) can be represented
by the surface of a Poincar{\'e} sphere with the horizontal
$|\textrm{H}\rangle$ (even, $|\textrm{e}\rangle$) and vertical
$|\textrm{V}\rangle$ (odd, $|\textrm{o}\rangle$) states located at
antipodes. (Any qubit can, of course, be represented by the surface
of a Poincar{\'e} sphere; consider, for example,
Ref.~\cite{Padgett99OL}, in which this treatment is applied to
classical light in first-order Gaussian modes after truncating all
other modes.) The Pauli operator $\sigma_{x}$ in polarization space
is a half-wave plate rotated by 45$^{\circ}$ with respect to
$|\textrm{H}\rangle$, while its pseudo-spin counterpart in parity
space is a simple phase plate that introduces a phase shift of $\pi$
between the two halves of the plane (a parity flipper, PF). The
Pauli operator $\sigma_{z}$ in polarization space is a half-wave
plate, while its pseudo-spin counterpart in parity space is a
spatial flipper (SF), a device that flips the beam in space
$\phi(x)\rightarrow\phi(-x)$, which may be implemented with a
mirror, for example.

With these building blocks in hand, we construct some operators
essential for quantum information processing. A rotation $\bf
R(\theta)$ in polarization space is implemented by a polarization
rotator, for example, and in parity space by a phase plate that
introduces a phase $\theta$ between the two half planes (a parity
rotator, PR). An even beam $|\textrm{e}\rangle$ incident on a phase
plate with a $\theta=\pi$ difference between the two halves (a PF)
obviously becomes odd $|\textrm{o}\rangle$, and vice versa. Less
obvious is the fact that introducing a phase difference
$\theta=\frac{\pi}{2}$ between the two halves of the plane of an
even mode transforms it into the equal superposition
$\frac{1}{\sqrt{2}}\{|\textrm{e}\rangle+i|\textrm{o}\rangle\}$.
Finally, a polarizing beam splitter that projects the polarization
state by separating the two orthogonal components into separate
spatial paths has its counterpart in a parity analyzer (PA): this
device is a balanced Mach--Zehnder interferometer (MZI)
\cite{Saleh07BOOK} with a SF placed in one arm that serves to
separate the $|\textrm{e}\rangle$ and $|\textrm{o}\rangle$
components of an incoming state into two separate spatial paths.
Thus quantum information processing experiments conducted on
photonic qubits in polarization space may be readily implemented in
1D spatial-parity space using Fig. 1 as a Rosetta stone that guides
the translation between these implementations. Previous work has
identified similar devices for manipulating spatial parity
\cite{Sasada03PRA}, but without identifying the underlying qubit
structure of the photon field.

\begin{figure}[ht]
\includegraphics[width=3.4in,height=2in]{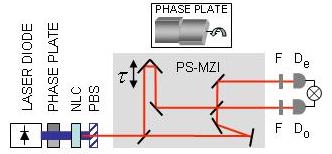}
\caption{(Color online) Schematic of the experimental arrangement.
NLC: nonlinear crystal; PBS: polarizing beam splitter; F:
interference filter; D: detector; $\otimes$: coincidence circuit.
The PS-MZI serves as a parity analyzer. The inset shows the
construction of the phase plate (parity rotator) placed in the path
of the pump. It comprises two glass microscope slides, abutted at
the center of the pump beam, that can be tilted with respect to each
other.} \label{Setup}
\end{figure}

In this Letter, we demonstrate the ability to control, in a precise
manner, the generation of entangled two-photon states in parity
space by manipulating a classical parameter: the pump spatial
parity. The experimental arrangement is shown schematically in
Fig.~\ref{Setup}(a). A linearly polarized monochromatic pump laser
diode (wavelength 405 nm, power 50 mW) with an even spatial profile
illuminates a 1.5-mm-thick $\beta$-barium borate (BBO) nonlinear
optical crystal (NLC) in a collinear type-I configuration (signal
and idler photons have the same polarization, orthogonal to that of
the pump), after passing through a phase plate that serves as a PR
for the pump spatial parity. The pump is removed using a polarizing
beam splitter placed after the crystal as well as by interference
filters (centered at 810 nm, 10-nm bandwidth) placed in front of the
detectors $\rm{D_e}$ and $\rm{D_o}$ (EG$\&$G SPCM-AQR-15-FC), the
outputs of which are fed to a coincidence circuit (denoted
$\otimes$) and thence to a counter. The signal and idler photons are
directed to a parity-sensitive MZI (PS-MZI), which, at a relative
path delay $\tau=0$, serves as a PA, as described above. For
purposes of comparison, we also carry out all the experiments with
the SF removed, corresponding to a traditional MZI.

It can be shown \cite{Abouraddy07PRA} that an even (odd) pump
results in downconverted photonic qubits in a $|\Phi^{+}\rangle$
($|\Psi^{+}\rangle$) parity state,
\begin{eqnarray}
|\textrm{even}\rangle_{\textrm{p}}&\rightarrow&|\textrm{even}\rangle_{\textrm{s}}|\textrm{even}\rangle_{\textrm{i}}+\,|\textrm{odd}\rangle_{\textrm{s}}|\textrm{odd}\rangle_{\textrm{i}}\nonumber\,=|\Phi^{+}\rangle,\\
|\textrm{odd}\rangle_{\textrm{p}}&\rightarrow&|\textrm{even}\rangle_{\textrm{s}}|\textrm{odd}\rangle_{\textrm{i}}\,\,+|\textrm{odd}\rangle_{\textrm{s}}|\textrm{even}\rangle_{\textrm{i}}=|\Psi^{+}\rangle,
\end{eqnarray}
where p, s, and i refer to the pump, signal, and idler,
respectively. Placing the PR in the pump beam results in the
superposition
$\cos\theta|\textrm{even}\rangle_{\textrm{p}}+i\sin\theta|\textrm{odd}\rangle_{\textrm{p}}$,
where varying $\theta$ traces out the great circle, indicated by the
dashed arrow, on the surface of the Poincar{\'e} sphere in Fig.
\ref{Comparison}. The SPDC photons produced by such a pump are
generated in the superposition state
$\cos\theta|\Phi^{+}\rangle+i\sin\theta|\Psi^{+}\rangle$. An
interesting consequence of controlling the pump parity in this
fashion is that the two-photon state is guaranteed to remain
\textit{maximally entangled}; the concurrence $C$
\cite{Wootters97PRL} of the state is always maximal ($C=1$) for all
$\theta$. This is in contrast to using a pump in the superposition
$\cos\theta|\textrm{even}\rangle_{\textrm{p}}+\sin\theta|\textrm{odd}\rangle_{\textrm{p}}$,
where varying $\theta$ traces out a different great circle on the
surface of the Poincar{\'e} sphere, such that the concurrence
$C=|\cos2\theta|$, whereupon a \textit{separable} state ($C=0$) is
obtained when $\theta=\frac{\pi}{4}$.

Before presenting the experimental results, we provide a heuristic
overview of the interference effects expected in this arrangement.
The two entangled photons are incident at the same port of a beam
splitter; two distinct cases arise, leading to qualitatively
different interference patterns. In the first case, each photon
emerges from a different beam-splitter port. When these photons are
brought back at a second beam splitter, after a delay $\tau$ in one
of the arms (see Fig.~\ref{Setup}), the well-known Hong--Ou--Mandel
(HOM) \cite{HOM87PRL} dip is observed in the coincidence rate
$G^{(2)}(\tau)$. In the second case, the two photons emerge together
from either output port of the first beam splitter. For SPDC with a
monochromatic pump, the frequencies of the two photons are
anti-correlated so that $\omega_s=\frac{\omega_{p}}{2}+\Omega$ and
$\omega_i=\frac{\omega_{p}}{2}-\Omega$, where $\frac{\omega_p}{2}$
is half the pump frequency and $\Omega$ is a deviation therefrom. A
delay $\tau$ will then lead to a fixed phase difference
$\exp\{-i(\frac{\omega_{p}}{2}+\Omega)\tau\}\exp\{-i(\frac{\omega_{p}}{2}-\Omega)\tau\}=\exp\{-i\omega_{p}\tau\}$
between the two paths, whereupon $G^{(2)}(\tau)$ will simply be a
sinusoid at the pump period $\textsf{\textsl{T}}_{\textrm{pump}}$
\cite{Rarity90PRL}. These two cases coexist in the experimental
arrangement shown in Fig. \ref{Setup}, resulting in a coincidence
interferogram that combines the HOM dip and the sinusoid at the pump
period. Exact expressions for the two-photon interferogram that take
into account both the temporal and spatial aspects are readily
derived using the formalism in Refs. \cite{Saleh00PRA}.

\begin{figure}[ht]
\includegraphics[width=3.14in,height=1.3in]{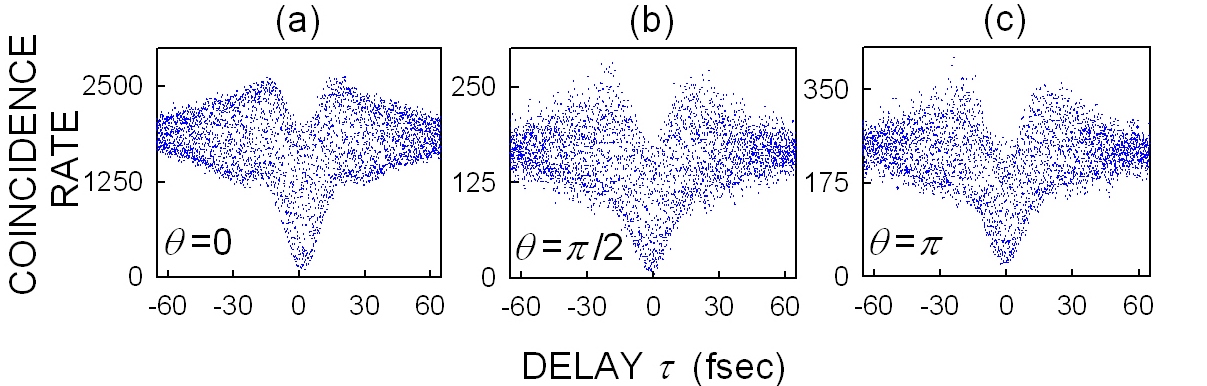}
\caption{(Color online) Observed coincidence rate from a traditional
MZI for three different settings of a parity rotator placed in the
pump: (a) $\theta=0$ (even pump), (b) $\theta=\frac{\pi}{2}$ (equal
superposition of even and odd), and (c) $\theta=\pi$ (odd pump).}
\label{Mach-Zehnder_Data}
\end{figure}

Since these interference effects exist independently of the spatial
parity, they may be observed using a traditional MZI
\cite{Rarity90PRL}. In Fig.~\ref{Mach-Zehnder_Data}, we present
coincidence rates for three different settings of a PR placed in the
pump beam (Fig.~\ref{Setup}). The three coincidence interferograms
displayed for the MZI are those of an even pump ($\theta=0$), an odd
pump ($\theta=\pi$), and a pump in an equal superposition of even
and odd parity ($\theta=\frac{\pi}{2}$). The coincidence rate as a
function of $\tau$ exhibits the two aforementioned features: an HOM
dip (whose width is inversely related to the SPDC bandwidth) and a
sinusoid with the period of the pump laser. It is obvious from the
observed coincidence interferograms in Fig.~\ref{Mach-Zehnder_Data}
that the traditional MZI is oblivious to the spatial parity of the
incident light.

\begin{figure*}[ht]
\includegraphics[width=6.5in,height=2.125in]{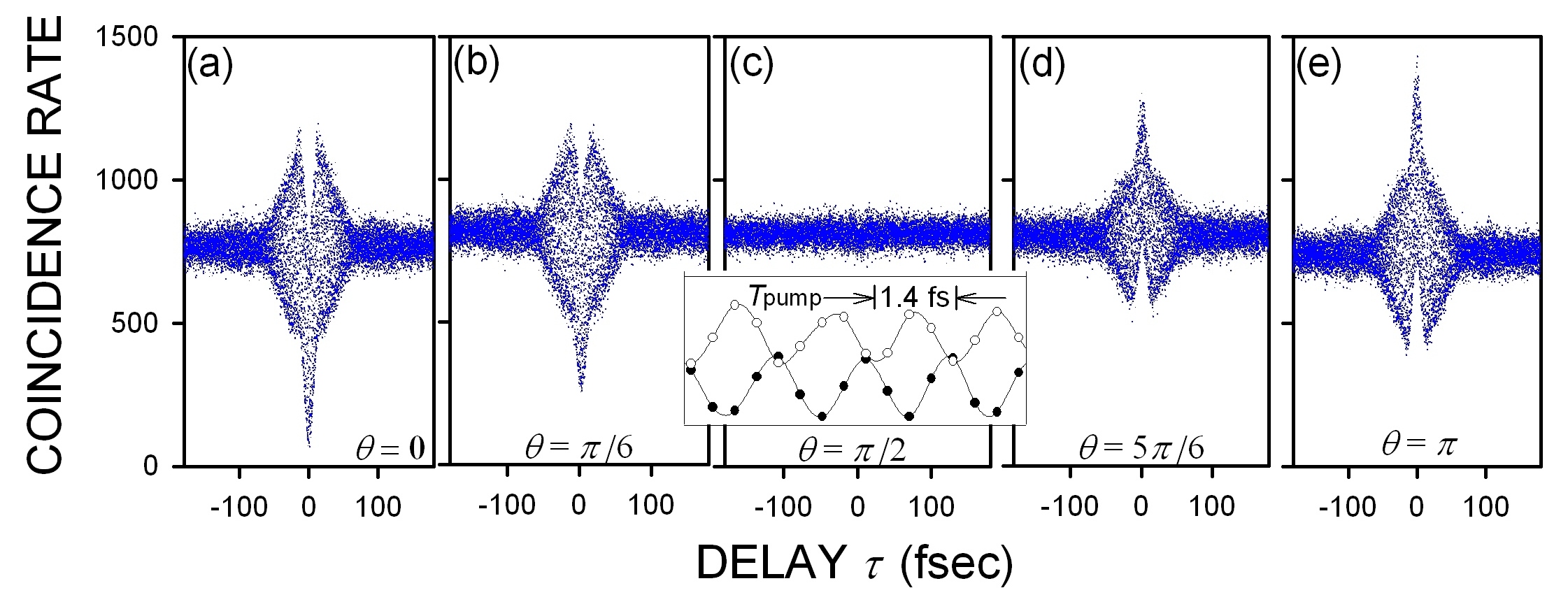}
\caption{(Color online) Coincidence rate at the output of a PS-MZI
for different settings of the parity rotator placed in the pump
beam. The five settings are (a) $\theta=0$ (pump is even), (b)
$\theta=\frac{\pi}{6}$, (c) $\theta=\frac{\pi}{2}$ (equal
superposition of even and odd), (d) $\theta=\frac{5\pi}{6}$, and (e)
$\theta=\pi$ (pump is odd). The inset shows oscillation at the pump
period $\textsf{\textsl{T}}_{\textrm{pump}}$ for panels (a) (full
circles) and (e) (open circles) near $\tau = 0$. The two curves,
which are spline fits, are 180$^{\circ}$ out of phase, as expected.}
\label{PA_Data}
\end{figure*}

The experimental results are altered dramatically when this
traditional MZI is converted into a PS-MZI by the insertion of an SF
in one of its arms. When the $|\Psi^{+}\rangle$ state is generated
(corresponding to an odd pump), the photons have opposite spatial
parity, and hence emerge from different output ports of the PA,
thereby producing a coincidence count. The two photons in the
$|\Phi^{+}\rangle$ state (corresponding to an even pump), on the
other hand, have the same parity and hence emerge together from the
same output port and do not produce a coincidence count. We expect
that the $|\Phi^{+}\rangle$ state will produce a minimum in the
coincidence rate at $\tau=0$, while the $|\Psi^{+}\rangle$ state
will produce a maximum. Both of these predictions are borne out by
the experimental results for the PS-MZI shown in Fig. \ref{PA_Data}.
The five panels correspond to different settings of the PR, varying
from $\theta=0$ to $\theta=\pi$. The high-visibility HOM \emph{dip}
at $\theta=0$ gradually loses visibility as $\theta$ increases,
resulting in a featureless interferogram at $\theta=\frac{\pi}{2}$.
Increasing $\theta$ further results in the emergence of an HOM
\emph{peak}, which attains its maximal visibility at $\theta=\pi$.

\begin{figure}[ht]
\includegraphics[width=3.15in,height=1.65in]{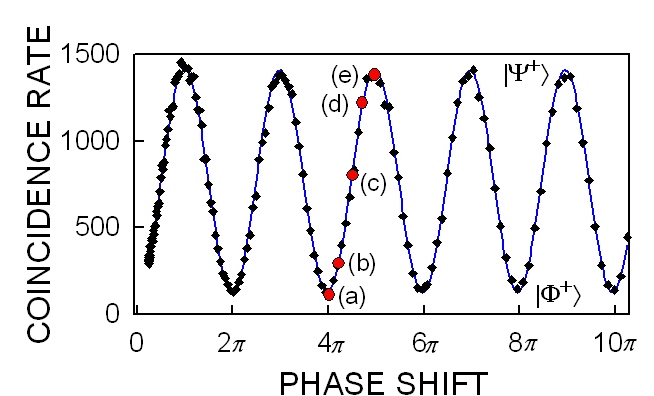}
\caption{(Color online) Coincidence rate at $\tau=0$ of the PS-MZI
as the parity of the pump is continuously varied over the range
$0<\theta<10\pi$. The diamonds represent experimental data; the
solid curve is a theoretical fit. The five circles labeled (a)--(e)
correspond to the five settings used in the panels of Fig.
\ref{PA_Data}.} \label{ParityRotatorInPump}
\end{figure}

To more explicitly demonstrate our ability to precisely control the
generation of the entangled SPDC state in parity space, we have
carried out an experiment in which we hold the delay fixed at
$\tau=0$ while varying the angle $\theta$ of the PR in the pump. The
coincidence rate shown in Fig. \ref{ParityRotatorInPump} oscillates
between maxima corresponding to the state $|\Psi^{+}\rangle$
produced at $\theta=\pi,3\pi,5\pi,...$ and minima at
$\theta=0,2\pi,4\pi,...$ corresponding to the state
$|\Phi^{+}\rangle$.  The curve thus obtained represents five
complete circumnavigations of the Poincar{\'e} sphere representing
the pump's spatial parity.  Confirmation of the presence of
entanglement is achieved by demonstrating a Bell-inequality
violation, as we have shown in Ref.~\cite{Yarnall07arXiv}.

In conclusion, we have demonstrated the controlled synthesis of
two-photon states entangled in the parity of one dimension of their
transverse spatial distribution. The prescribed states are
controlled by manipulating the spatial parity of the pump, a
classical parameter, and not by direct manipulation of the generated
entangled photons. Furthermore, we constructed a parity-sensitive
MZI by adding only one mirror to one arm of a traditional MZI, and
showed that when the arms have equal path lengths, this device acted
as a parity analyzer. This seemingly insignificant change
dramatically altered the behavior of the interferometer. It acquired
sensitivity to spatial parity, which a traditional MZI lacks, and
was used to analyze maximally entangled qubits in the spatial-parity
basis. Our approach is inherently interesting from the point-of-view
of quantum information processing. While each photon carries
\textit{one} polarization qubit, it has \emph{two} transverse
dimensions, and one qubit can be encoded in each. Furthermore, these
two spatial-parity qubits per photon may be led to interact using
simple optical arrangements. The two-photon SPDC state thus carries
four parity qubits, allowing the study of hyperentanglement
\cite{Barreiro05PRL} in a straightforward manner.

\begin{acknowledgments}\textit{Acknowledgments}---This work was supported by a U.~S.~Army
Research Office (ARO) Multidisciplinary University Research
Initiative (MURI) Grant and by the Center for Subsurface Sensing and
Imaging Systems (CenSSIS), an NSF Engineering Research Center. This
work is sponsored by the National Aeronautics and Space
Administration under Air Force Contract \#FA8721-05-C-0002.
Opinions, interpretations, recommendations and conclusions are those
of the authors and are not necessarily endorsed by the United States
Government. A.F.A. acknowledges the generous support and
encouragement of Y.~Fink and J.~D.~Joannopoulos.
\end{acknowledgments}

\end{document}